\documentclass[preprint,authoryear,12pt]{elsarticle}

\usepackage{graphicx}

\usepackage{amssymb}

\journal{Advances in Space Research}

\begin{document}

\begin{frontmatter}

\title{Probing the origin of the iron K$\alpha$ line around stellar and supermassive black holes using X-ray polarimetry}

\author{F.~Marin}
\ead{frederic.marin@astro.unistra.fr}
\author{F.~Tamborra}
\address{Observatoire Astronomique de Strasbourg, Universit\'e de Strasbourg, 
       CNRS, UMR 7550, 11 rue de l'Universit\'e, 67000 Strasbourg, France}

\begin{abstract}
Asymmetric, broad iron lines are a common feature in the X-ray spectra of both X-ray binaries (XRBs) and type-1 
Active Galactic Nuclei (AGN). It was suggested that the distortion of the Fe~K$\alpha$ emission results from Doppler 
and relativistic effects affecting the radiative transfer close to the strong gravitational well of the central 
compact object: a stellar mass black hole (BH) or neutron star (NS) in the case of XRBs, or a super massive black hole 
(SMBH) in the case of AGN. However, alternative approaches based on reprocessing and transmission of radiation 
through surrounding media also attempt to explain the line broadening. So far, spectroscopic and timing analyzes 
have not yet convinced the whole community to discriminate between the two scenarios. Here we study to which extent 
X-ray polarimetric measurements of black hole X-ray binaries (BHXRBs) and type-1 AGN could help to identify the possible 
origin of the line distortion. To do so, we report on recent simulations obtained for the two BH flavors and show that 
the proposed scenarios are found to behave differently in polarization degree and polarization angle. A relativistic origin 
for the distortion is found to be more probable in the context of BHXRBs, supporting the idea that the same mechanism 
should lead the way also for AGN. We show that the discriminating polarization signal could have been detectable by 
several X-ray polarimetry missions proposed in the past.
\end{abstract}


\end{frontmatter}

\parindent=0.5 cm

\section{Introduction}
\label{Intro}

The X-ray spectrum observed in XRBs and AGN is complex and constituted by several components, due to the circumnuclear material which 
scatters, absorbs and reprocesses photons produced by the disk \citep{Risaliti2004}. In AGN, the accretion disk produces thermal photons 
(the so called ``multi-temperature black body emission''), whose spectral energy distribution peaks in the UV band for a 
$\sim 10^8$~M$_\odot$ SMBH \citep{Pringle1972,Shakura1973}. In XRBs, the thermal emission produced by a BH with $\sim 10$~M$_\odot$ peaks in 
the soft X-ray band (i.e. a few keV, $ibid.$). To produce the hard X-ray observed in both cases, an optically thin, hot corona of thermally 
distributed electrons is believed to Comptonize (i.e. via the inverse Compton effect) soft photons to higher energies 
\citep{Haardt1991,Haardt1993}. The higher energetic component of the X-ray radiation that falls back to the disk is reflected 
while the less energetic part ($\lesssim 10$~keV) is absorbed and partly reprocessed in emission lines. The combination of 
the relatively strong abundance of iron in the disk matter and the associated high Fe fluorescence yield makes the Fe K-shell 
fluorescence lines the strongest emission features to be detected in the X-ray wave band. They naturally became one of the 
prior targets of X-ray observations, as iron lines can be used as probes of matter under extreme conditions \citep{Nandra1997,Reynolds1997}. 

Exploring the accretion disks physics in AGN and X-ray binaries, \citet{Fabian1989} predicted that the iron emission lines 
should be distorted by Doppler and general relativistic effects acting close to the central BH. The \textit{ASCA}/\textit{SIS} 
observation of the Seyfert~1 galaxy MCG-6-30-15 confirmed such hypothesis, farther strengthened by subsequent systematic 
surveys of radio-quiet, type-1 AGN, where broadened and distorted iron lines where found in a few tens of targets 
\citep{Reeves2006,Nandra2007,delaCalle2010,Patrick2011,Patrick2012}. But what about stellar-mass compact objects? 
Back in the 1990's, \textit{ASCA}'s sensitivity and resolution limits prevented to clearly detect a broad red wing of 
iron lines in black hole X-ray binaries. In the case of neutron stars, the weaker iron line intensities made the detection 
even less probable. With the launch of \textit{XMM-Newton} and later \textit{Suzaku}, the detection of distorted Fe~K$\alpha$ 
lines then extended to stellar-mass black holes in binary systems \citep{Miller2004} and neutron star XRBs \citep{Bhattacharyya2007}, 
although \citet{Done2010} and \citet{Ng2010} challenged this view pointing out potential problems due to instrumental effects and/or
uncertainties in the spectroscopic analysis.

The distorted Fe~K$\alpha$ emission line at 6.4~keV is thus a feature shared by stellar and supermassive black hole powered objects 
\citep{Nandra2007,Miller2007,Cackett2008}. It seems reasonable to assume that the red-wing extension is due to the same broadening 
mechanisms: Doppler and general relativity effects. However, taking into account the interaction between radiation and the hot coronal 
plasma above the accretion disk, or cold, distant obscuring material along the observer's line-of-sight, some thought about a different 
explanation in which the apparent red-wing is actually ascribed to the continuum emission and the line is not really broad 
\citep[e.g.][]{Inoue2003,Miller2008}. Relativistic scenarios usually point into the direction of rapidly spinning BH, while complex 
absorption in AGN systematically lowers the estimated BH spin as the line is much less related to the innermost stable circular orbit (ISCO) 
radius. In the case of low-mass XRBs (LMXRBs), Comptonization by the corona can cause overestimations of the equivalent width of the 
line and thus also lower the BH spin estimation \citep{Makishima1996,Ng2010}.

While actual spectral and timing analyses have not yet convinced the whole community regarding the preponderant mechanism responsible 
for the broad Fe~K$\alpha$ line, it is the scope of this paper to present a different and independent path: X-ray polarimetry. 
By adding two more, independent observables to the spectroscopic information, i.e. the polarization degree and the polarization 
position angle, we show how X-ray polarimetry could help to independently solve the issue.

\section{Overview of different broadening mechanisms}
\label{Overview}
The mechanism responsible for the broadening of the iron K$\alpha$ line is still matter of debate. In the following subsections, 
we summarize the basic concepts behind the common interpretations of line broadening. The various mechanisms actually do not exclude 
each other and possibly all of them contribute together in distorting the line profile. Nonetheless the scientific community still 
argues about which mechanism is predominant and which can be neglected. For a detailed and complete dissertation on the iron line 
studies we address the reader to the comprehensive review by \citet{Reynolds2003} and references therein.

\subsection{Relativistic reflection}
\label{Relat}
A characteristic double-peaked profile is common for broad iron lines. The spectroscopic split between the two peaks depends on 
the observer's viewing angle and is attributed to Doppler shifting due to the orbital motion of the reprocessing matter in 
the accretion disk. Due to special relativistic aberration, emission from the matter on the approaching portion of the orbit (commonly 
referred to as the blue peak) is boosted and enhances the blue peak intensity with respect to the red one. In addition to the Doppler 
and boosting effects, the complete line profile is red-shifted and broadened due to the gravitational potential and the transverse 
Doppler effect (i.e. special relativistic time dilation). For a disk seen almost face on, the latter redshift effects dominate, 
while for larger inclinations, Doppler effects are dominant. As a direct consequence, from the broadening of the line, in particular 
from the extension of the red tail to low energies, the spin of the black hole or the radius of a NS can be inferred 
\citep{Fabian1989,Laor1991,Dovciak2004,Brenneman2006,Dauser2013}. 

\subsection{Compton scattering broadening in XRBs}
\label{Compton}
To produce high energy X-ray photons, Comptonization (i.e. Inverse Compton scattering) by a hot corona of thermally distributed 
electrons is invoked. In AGN the thermal energy of these photons can be as large as hundreds of keV. In XRBs, a layer (a "corona") 
of electrons with a thermal energy of a few keV should arise. In this case, the 6.4~keV iron line photons can lose energy by Compton 
scattering with less energetic electrons. In the multiple scattering regime (i.e. for an optically thick corona) this can result in a 
red shifting of the line centroid and an asymmetric broadening (but much less than for the relativistic reflection) of the profile 
toward lower energies, mimicking the effect produced by relativistic effects. The Compton broadening is believed to be dominant for 
particular systems like LMXRBs with a NS as the collapsed object \citep{Ng2010}. Nonetheless, for any accreting source, a certain 
amount of broadening should be ascribed to simple Compton scattering of photons traveling through the corona.

\subsection{Complex absorption in AGN}
\label{Abso}
Exploring the inner core of AGN is rather challenging as several opaque media lie in the vicinity of the nucleus. While many XRBs 
spectra can be considered as relatively absorption-free, the presence of obscuring circumnuclear matter around the SMBH, as well 
as outflows complicate the picture by adding possible sources of absorption along the observer's line-of-sight. Taking into 
account the action of cold matter onto AGN spectra, another scenario explaining the asymmetrical broadening of the Fe~K$\alpha$ 
line emerged a decade ago \citep{Inoue2003}. In this prescription, the fluorescent emission is neither intrinsically broad, nor 
predominantly blurred by gravitational effects, but the spectral line shape is rather "carved out" by absorption in more distant, 
absorbing cloudlets. A distribution of optically thick media, located along the observer's line-of-sight and partially covering 
the primary X-ray source is ultimately responsible for both the flux variability and the apparent broadening of the iron line. 
The overall extended red wing reduces to the sum of the uncovered continuum radiation, the transmitted and scattered radiation 
escaping "through the holes between the clouds", and the absorbed flux \citep{Miller2008,Miller2009,Miller2013}.

\section{Spectropolarimetric modeling}
\label{Modeling}
The accuracy of spectroscopic fitting, and thus the constraint on the BH spin, depends on a precise distinction between the shape 
of the underlying continuum and the iron line. So far, various models presented claim to be able to reproduce the spectral shape of 
the intensity flux and sometimes also the observed time-dependency. In this context, we discuss the recent results obtained by our 
group that follow the independent path of X-ray polarimetry. We illustrate the constraining and discriminating power of X-ray 
polarimetry by exploring the polarization signatures of different proposed scenarios. It is not the scope of this paper to produce
accurate spectral fits. We rather rely on the prescriptions given by certain models assuming reflection \citep[e.g.][]{Miniutti2004}, 
Comptonization \citep[e.g.][]{Pozdnyakov1983,Sunyaev1984,Hirano1987,Haardt1994} and absorption \citep[e.g.][]{Miller2008,Miller2009}.

\subsection{LMXRBs: impact of Compton scattering on the Fe K$\alpha$ line}
\label{LMXRBs}
To test the impact of broadening due to Compton scattering, we used {\sc MoCA}, a Monte Carlo code devoted to the study of spectra 
and polarization in accreting sources (Tamborra, Matt \& Bianchi, private communication). The code is written in IDL, a vectorized and interactive 
language. The two main features of the code are its modularity (with minor modifications, {\sc MoCA} can be applied to different 
accreting sources) and its full special relativistic treatment, using the Klein Nishina cross-section for the treatment of 
Compton scattering \citep{Klein1929}. MoCA belongs to the class of source-to-observer codes; it samples and follows each photon 
during its journey from the emitting source to the detector (the observer), saving the radiation's energy, direction, number of 
scattering events and the two Stokes parameters, Q and U, which describe the linear polarization degree and polarization angle. 
In this very first application, we do not use the full potential of the code yet, but we simply evaluate the effect of pure 
Compton scattering on line broadening. We want to offer an independent way to understand if the broad lines observed in NS 
are distorted by relativistic effects \citep{Cackett2008, Disalvo2009} or if the large broadening is overestimated by instrumental 
effects not properly taken into account (pile-up), thus reducing the impact of Compton broadening \citep{Ng2010}.

~\

According to a simple emissivity law $F~\propto~R^{-2.5}$ ($F$ being the flux and $R$ the disk radius), we generated monochromatic 
photons with energy 6.4~keV, arising isotropically from a non-rotating accretion disk with inner radius $R_{\rm i}$~=~6~$R_{\rm G}$ 
(gravitational radii, $R_{\rm G} = GM/c^2$ for black hole mass $M$) and outer radius $R_{\rm o}$~=~48~$R_{\rm G}$. Radiation travels 
trough a coronal region filled with free electrons of thermal energy kT$_{\rm e}$~=~2~keV. We considered two geometries for the corona: 
a spherical plasma cloud covering the inner half of the accretion disk (inner and outer radii of the corona being 
$R_{\rm c,i}$~=~6~$R_{\rm G}$ and $R_{\rm c,o}$~=~24 $R_{\rm G}$, respectively) and a thin slab gas covering the whole disk 
(height and outer radius of the slab being H~=~6~$R_{\rm G}$ and $R_{\rm c,o}$~=~48~$R_{\rm G}$, respectively). For both geometries, 
we computed an optically thin (Compton optical depth $\tau_{\rm c}$~=~0.1) and an optically thick corona ($\tau_{\rm c}$~=~1), 
covering a total of four scenarios. For each case, we generated $\sim 10^8$ photons.

\begin{figure}
   \begin{center}
   \includegraphics[width=12cm]{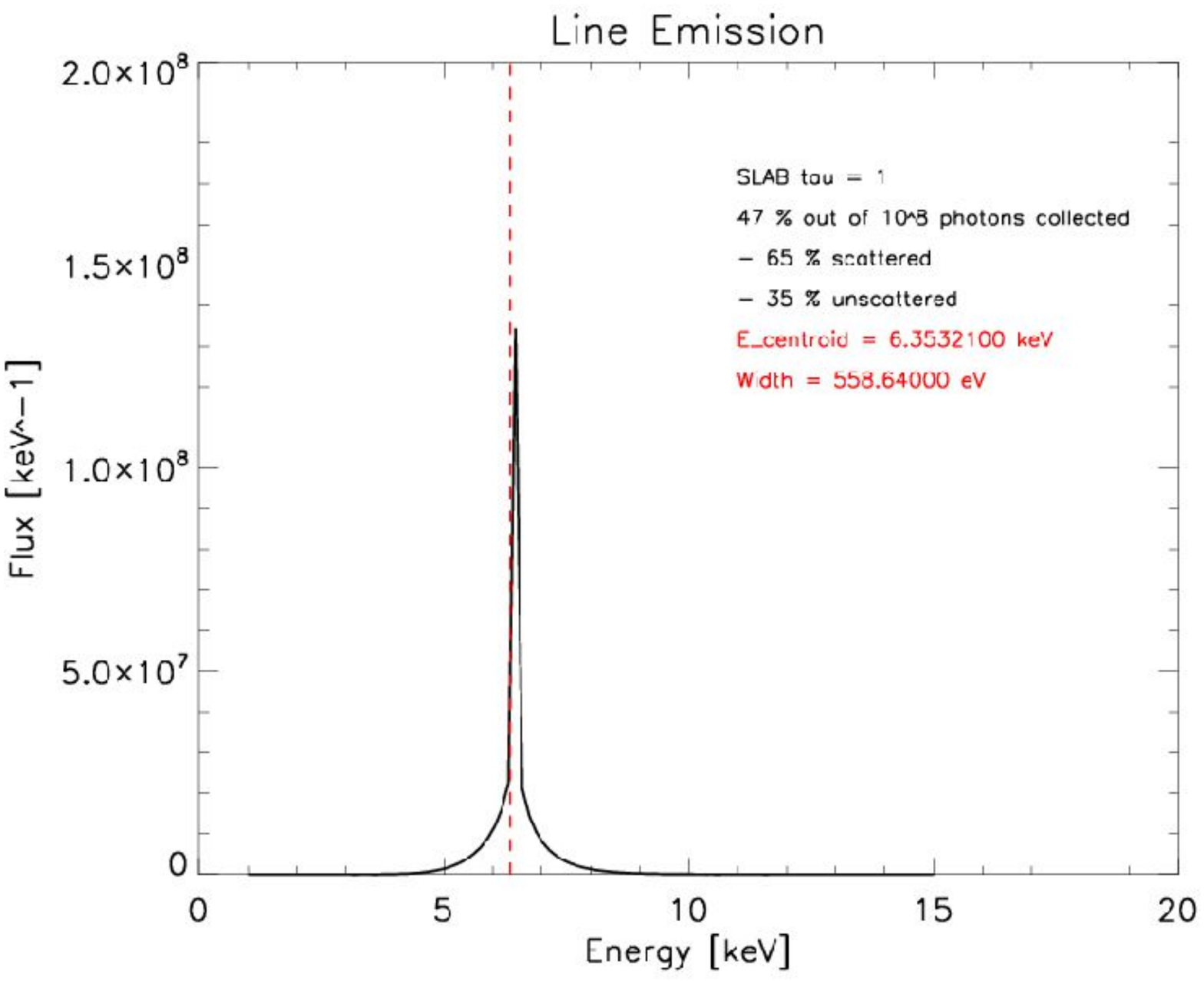}
   \includegraphics[width=12cm]{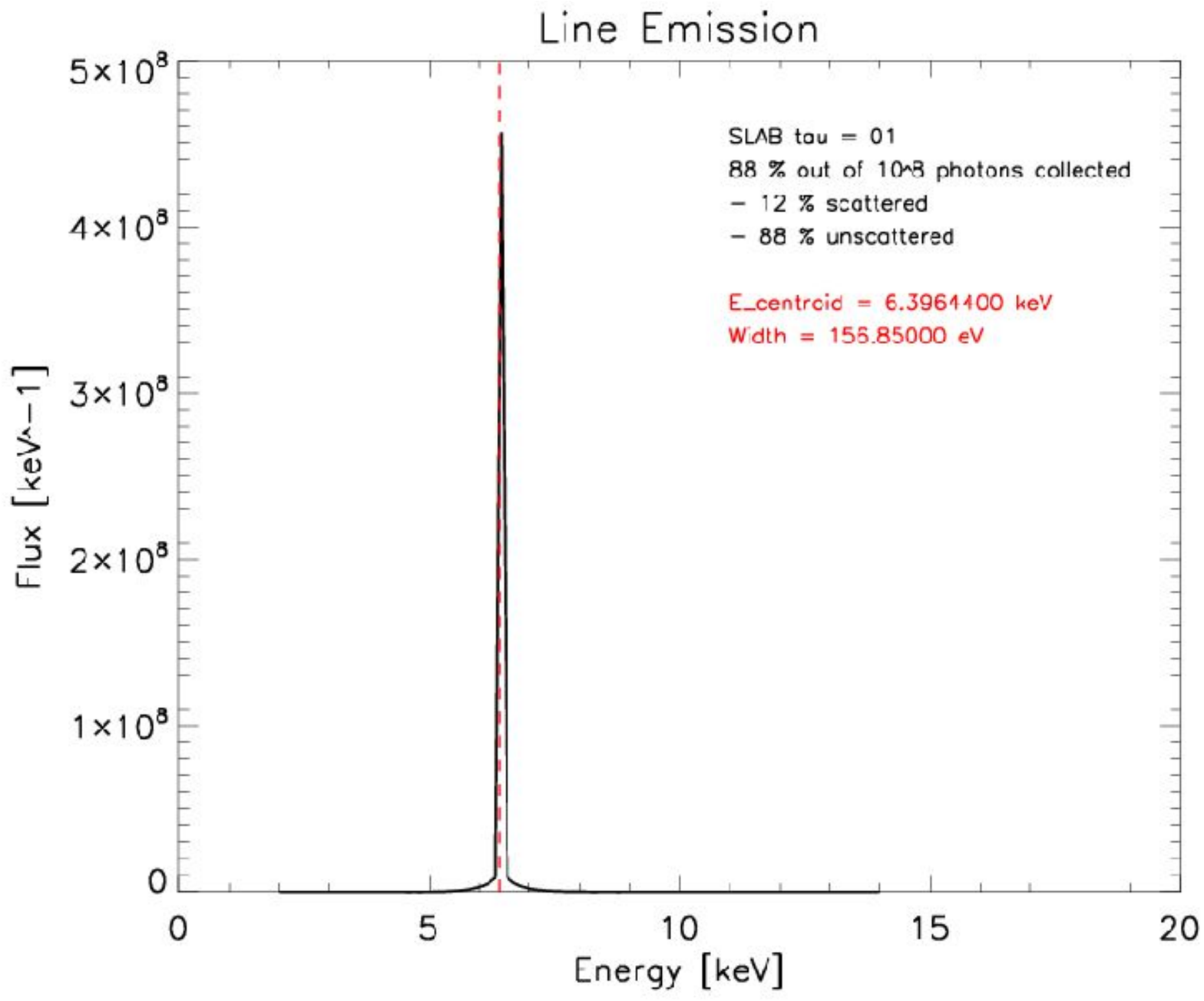} 
   \end{center}
   \caption{Intensity spectra (counts $\times$ energy) for the slab geometry, for an optically thick (top) and an optically thin (bottom) corona. 
	   The percentage of radiation that reached the observer (out of the initial $10^8$ photons emitted by the disk) is shown. 
	   We indicated also the sub-percentage of photons that experienced at least one scattering. The line centroid is 
	   indicated by the red-dashed line.}
   \label{Slab_geom_spectra}
\end{figure}

\begin{figure}
   \begin{center}
   \includegraphics[width=12cm]{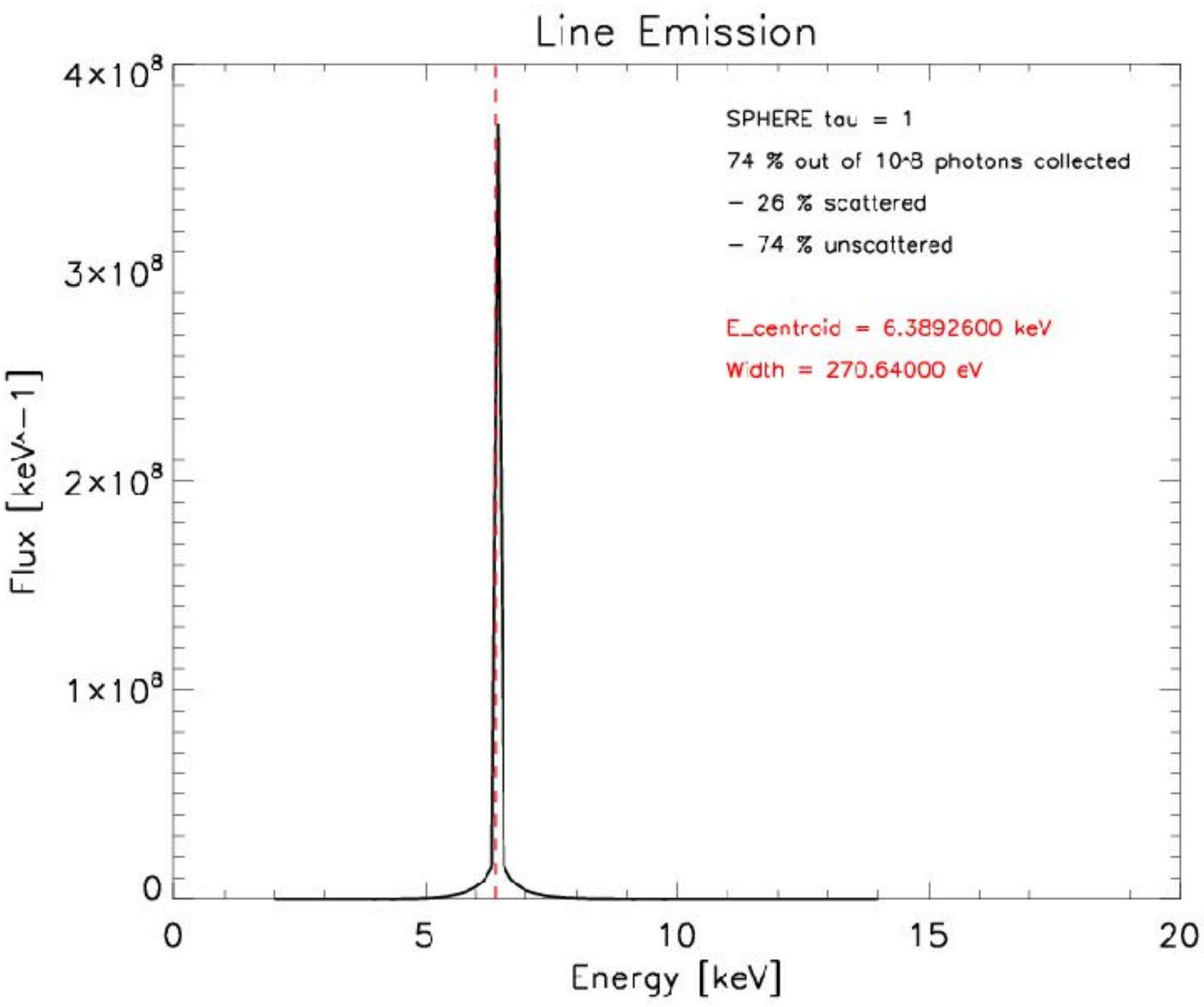} 
   \includegraphics[width=12cm]{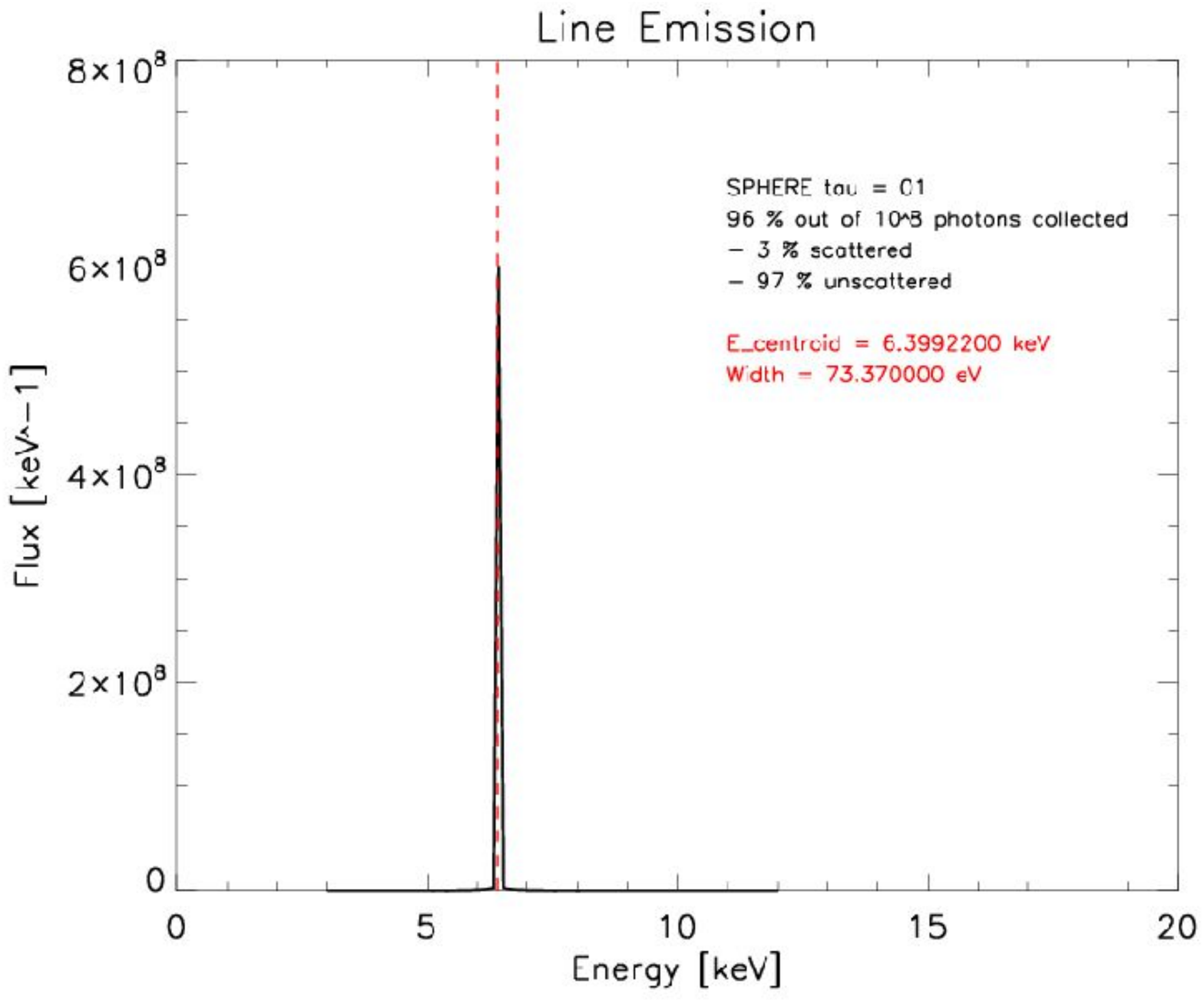} 
   \end{center}
   \caption{Intensity spectra (counts $\times$ energy) for the spherical geometry, for an optically thick (top) and an optically thin (bottom) corona. 
	  The percentage of radiation that reached the observer (out of the initial $10^8$ photons emitted by the disk) is shown. 
	  We indicated also the sub-percentage of photons who experienced at least one scattering. The line centroid is 
	  indicated by the red-dashed line.}
   \label{Sphere_geom_spectra}
\end{figure}

\begin{table}
    \begin{center}
\begin{tabular}{lll}
 & SLAB & SPHERE \\
 \hline
$\tau_{\rm c}$ = 1 & E$_c$ = 6.353 keV & E$_c$ = 6.389 keV\\
 & Width = 559 eV & Width = 271 eV \\
~\\ 
$\tau_{\rm c}$ = 0.1 & E$_c$ = 6.396 keV & E$_c$ = 6.399 keV\\
 & Width = 157 eV & Width = 73 eV
 \end{tabular}
 \caption{Line centroid and width for the slab (left column) and the spherical geometry (right column), 
	  for optically thick (first row) and optically thin (second row) coronas. The width of the line 
	  represents an upper limit to the equivalent width.}
  \label{Tab_line}
    \end{center}
\end{table}

\subsubsection{Spectral broadening}
\label{LMXRB:spectra}

In Fig.~\ref{Slab_geom_spectra} and Fig.~\ref{Sphere_geom_spectra}, we present the four intensity spectra produced by the 
two geometries in both of the optical depth regimes. According to the energy of the line centroid, summarized in Tab.~\ref{Tab_line}, 
the slab corona is found to be more efficient in terms of down-scattering, producing broader lines than the spherical plasma cloud. 
The related line widths (Tab.~\ref{Tab_line}) represent an upper limit of the real equivalent width, as the monochromatic radiation 
has been produced by the disk without any continuum emission. In order to compare spectroscopic simulations results with equivalent 
widths derived from observations analysis, a more complex model in which the line is produced together with the continuum is 
required, and will be provided in future work. Nonetheless, these qualitative results suggest that the $\sim$100~eV broadening 
produced by an optically thin corona is not large enough to solely associate the broadening mechanism with Compton scattering. 
Moreover, for the optically thick slab corona, the broadened line is found to be slightly asymmetric, and its centroid is red-shifted
by more than $1\%$ with respect to the initial 6.4~keV emission energy. The spherical corona, on the other hand, seems to be unable 
to efficiently shift the emission line at any of the two optical depths.

\subsubsection{Coronal polarization}
\label{LMXRB:polarization}

\begin{figure}
   \begin{center}
   \includegraphics[width=12cm]{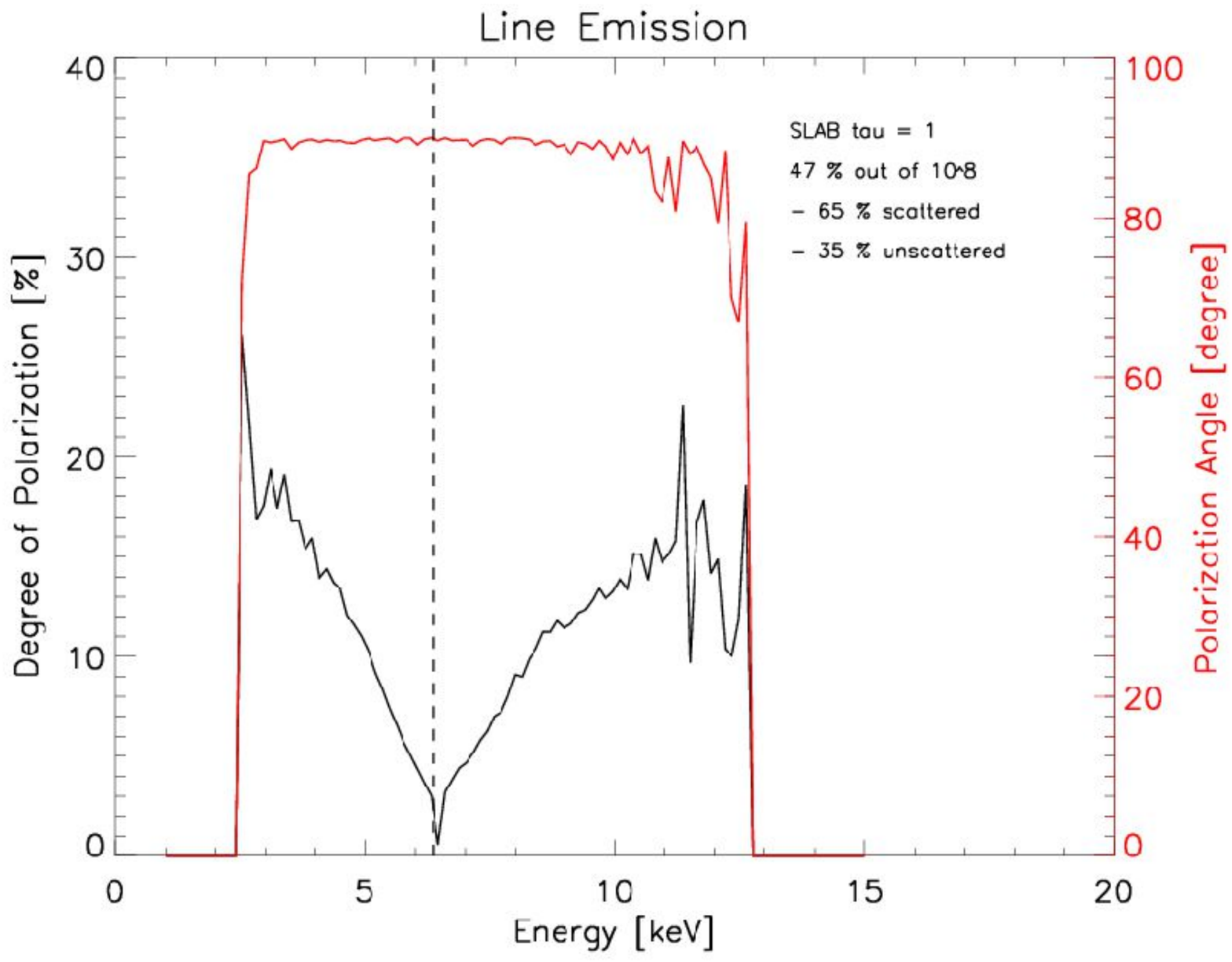} 
   \includegraphics[width=12cm]{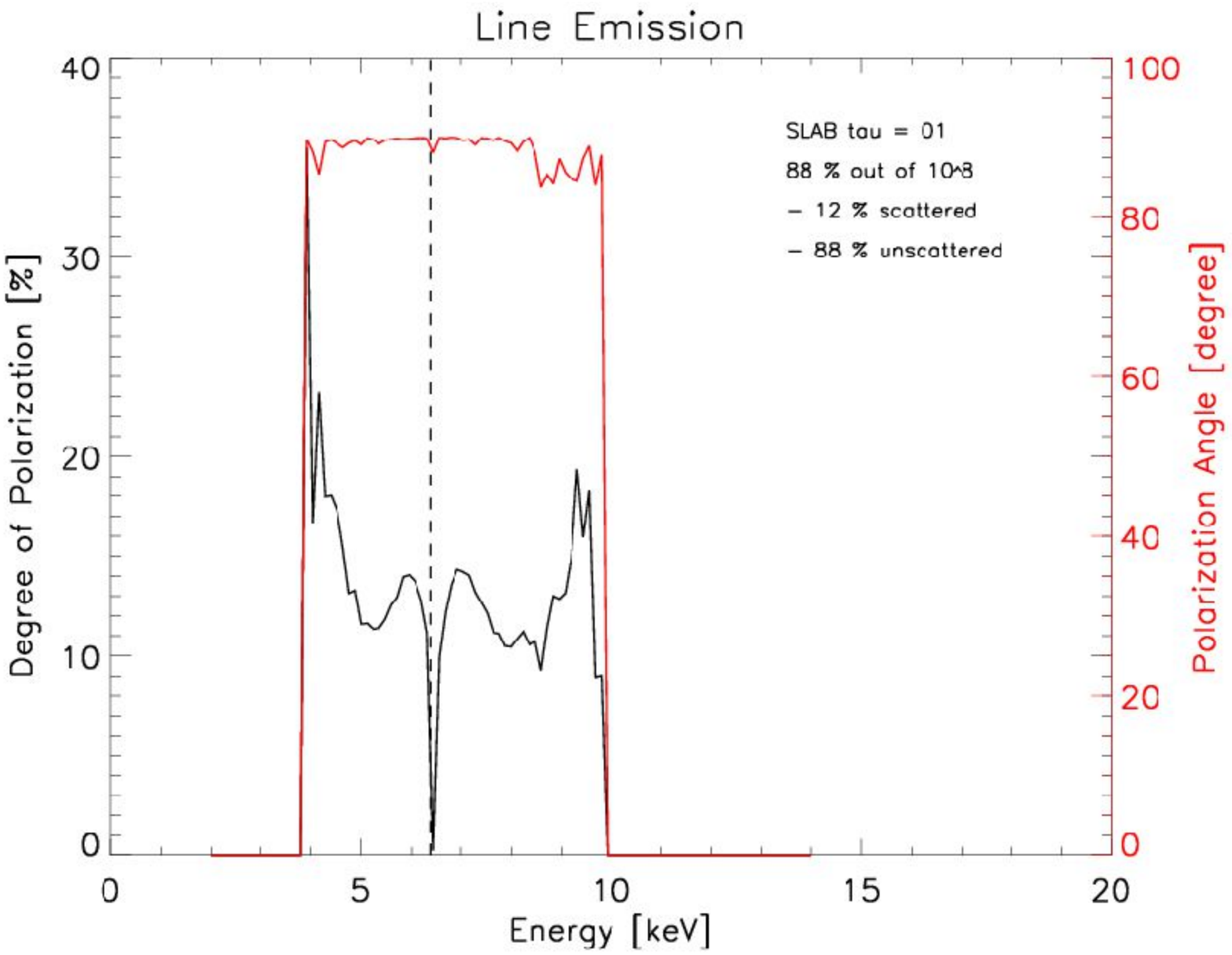} 
   \end{center}
   \caption{Polarization degree $P$ (black line) and corresponding polarization position angle $\Psi$ (red line) 
	  for the same slab coronas as in Fig.~\ref{Slab_geom_spectra}.}
   \label{Slab_geom_pol}
\end{figure}

\begin{figure}
   \begin{center}
   \includegraphics[width=12cm]{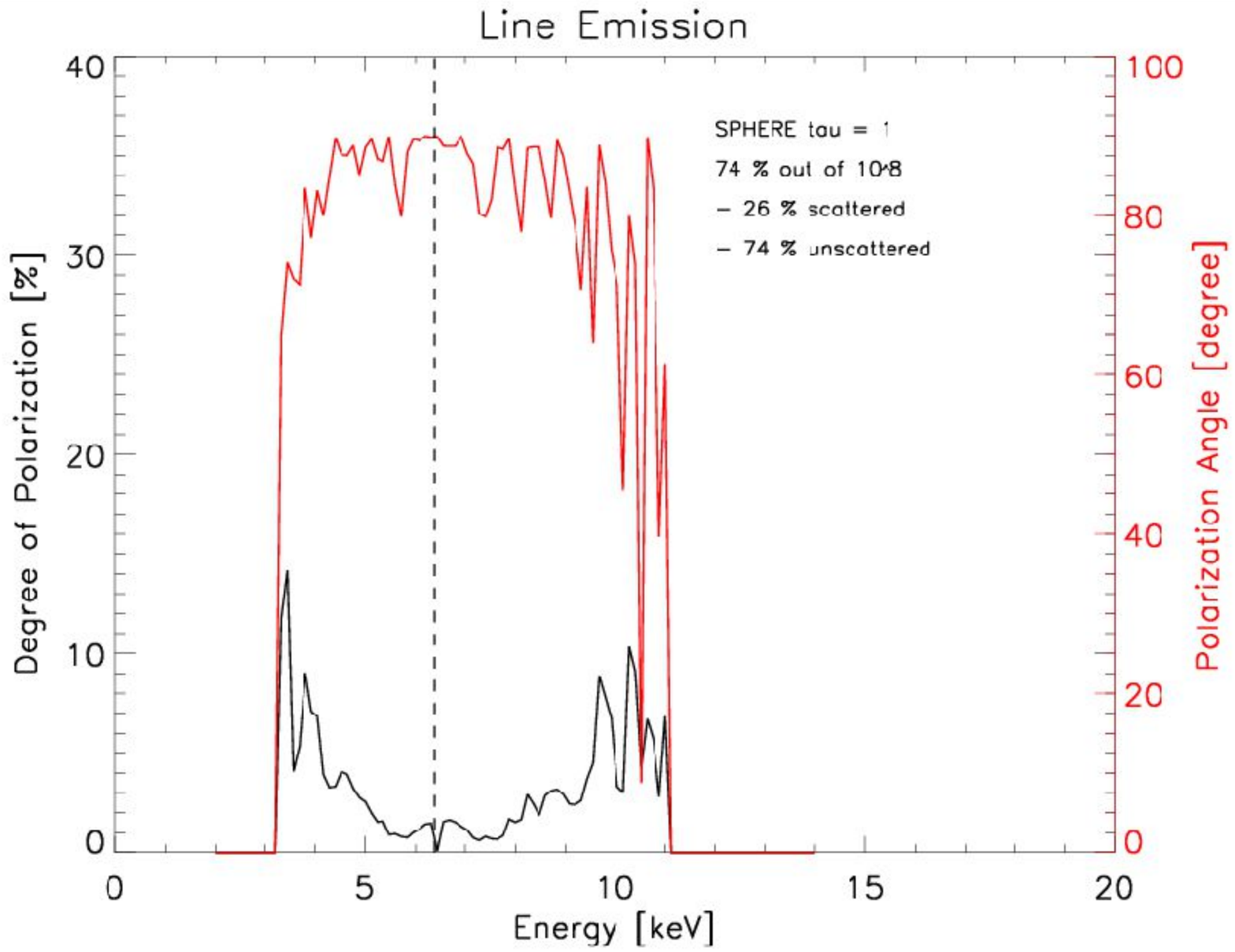} 
   \includegraphics[width=12cm]{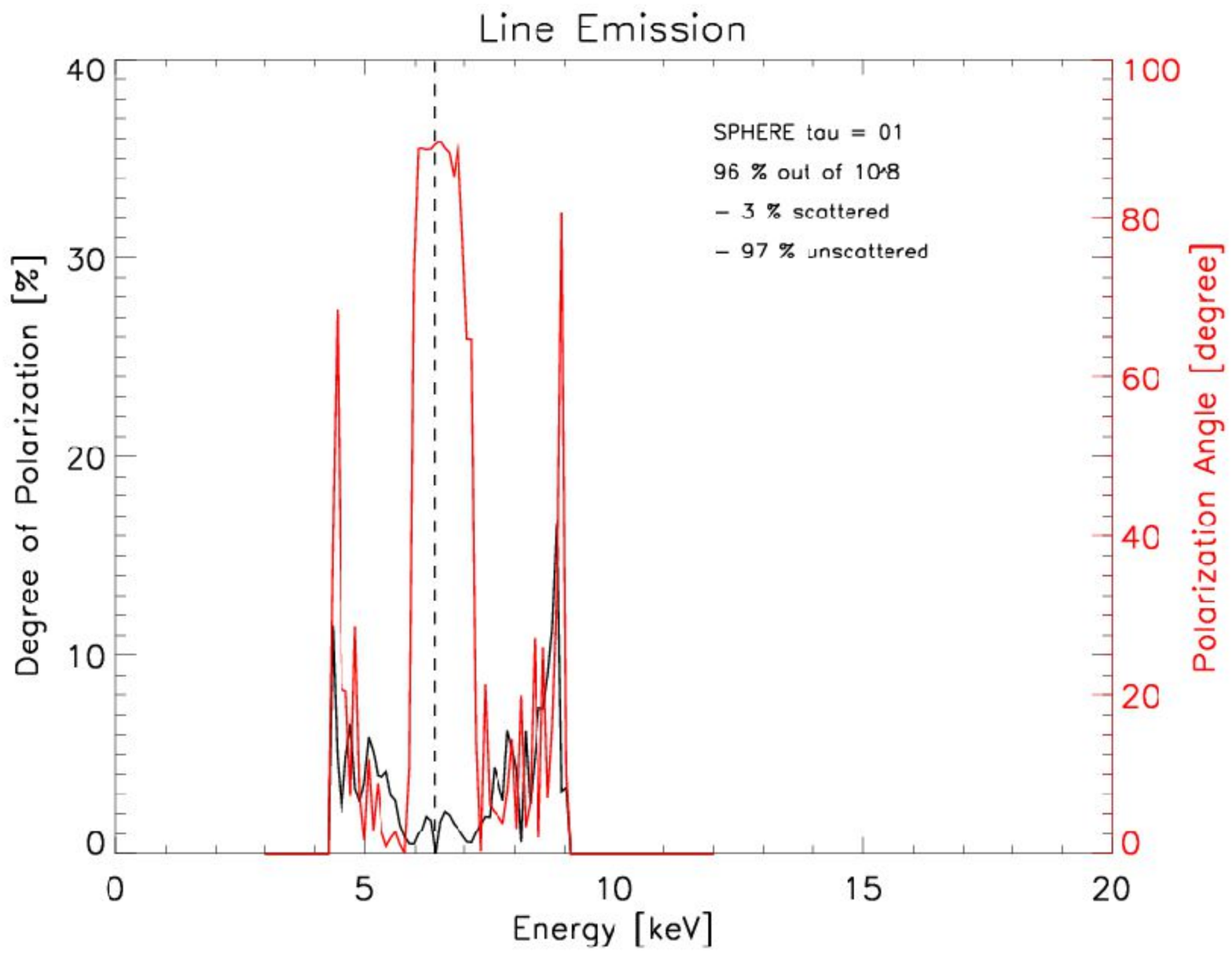} 
   \end{center}
   \caption{Polarization degree $P$ (black line) and corresponding photon polarization angle $\Psi$ (red line) 
	  for the same spherical coronas as in Fig.~\ref{Sphere_geom_spectra}.}
   \label{Sphere_geom_pol}
\end{figure}

To push our investigation of the impact of Comptonization onto the iron line broadening somewhat further, we present in Fig.~\ref{Slab_geom_pol} 
and Fig.~\ref{Sphere_geom_pol} the linear polarization degree $P$ (in percentage, averaged over all disk inclination) and the 
corresponding photon polarization angle $\Psi$ (in degrees) as a function of energy. By convention, a photon polarization angle of $0^{\circ}$ 
indicates that the photon's $\vec E$-vector is perpendicular to the projected disk axis, while $\Psi$ = $90^{\circ}$ means that the $\vec E$-vector 
is aligned with the disk axis. 
When the polarization degree is zero, the polarization angle has no physical meaning as it is a random superposition of all the incident photon's 
polarization planes. Monochromatic seed photons, thermally produced by the disk, are unpolarized, hence the resulting polarization signal is 
uniquely provided by photons which have experienced at least one scattering event before reaching the observer. There is a proportionality 
between the radiation energy shift, the number of scatterings and the final polarization degree. The noise seen in the polarization spectra 
is due to the Poissonian statistics of the Monte Carlo method and must not be taken into account for physical interpretation. Finally, we decided 
to set both $P$ and $\Psi$ to zero if the number of photons recorded by individual energy bin was less than 100 and thus negligible.

~\

The morphology plays a significant role for the polarization signal, as the anisotropy of the scattering geometry enhances the resulting polarization. 
Thus, a spherical corona produces more modest linear polarization percentages than a slab geometry. For both optical depths, a spherical morphology shows 
$P \le$ 10~\%, while the polarization degree induced by slab coronas is found to be larger than 10~\%. On each side of the emission line centroid, a 
bump in polarization degree appears. These features are due to the contribution of photons that experienced only one scattering event: the energy shift 
induced by single scattering is small but the polarization of the radiation increases. A second scattering event will shift the photons farther in 
energy and randomize the polarization vectors, thus decreasing $P$ and resulting in the polarization bumps just left and right of the almost unpolarized 
line centroid as shown in Fig.~\ref{Slab_geom_pol} and Fig.~\ref{Sphere_geom_pol}. 

~\

Additional information can be extracted from the polarization angle $\Psi$ (plotted in red in Fig.~\ref{Slab_geom_pol} and Fig.~\ref{Sphere_geom_pol}). 
The slab corona, Fig.~\ref{Slab_geom_pol}, produces a net polarization position angle aligned with the projected slab symmetry axis, independently 
of the optical thickness. Since the last scattering occurs mainly in a plane oriented roughly perpendicularly to the disk axis, the resulting polarization 
angle remains predominantly parallel. Even for the case of multiple reprocessing in an optically thick disk, an energy-independent, parallel
$\Psi$ is obtained, which confirms previous simulations in \citet{Goosmann2007} and \citet{Marin2012a}.

The geometry of a spherical corona (Fig.~\ref{Sphere_geom_pol}) results in a different behavior of $\Psi$ as a function of the optical thickness. 
In the optically thick scenario, the first scattering event occurs close to the disk surface and the photon's energy is only slightly shifted. So, 
between 5 and 7~keV, the corona behaves similarly to a slab geometry, and the net polarization position angle is 90$^\circ$. Secondary and tertiary 
scatterings modify the radiation's energy, shifting it below 5~keV or higher than 7~keV, and randomizing $\Psi$, as photons have traveled farther 
through the symmetric geometry of the sphere. The resulting photon position angle becomes random. The optically thin prescription 
(Fig.~\ref{Sphere_geom_pol}, bottom) presents a switch of the polarization position angle, from 90$^\circ$ to 0$^\circ$, at energies corresponding 
to the $P$ bumps. It indicates that the main contribution to polarization is given by singly scattered, parallelly polarized photons. The photon 
position angle becomes equal to 0$^\circ$ beyond the polarization bumps as there is no contribution of multiple scattering (the corona being optically thin),
and $\Psi$ is given by the few recorded photons with a substantially random polarization position angle.

~\

Using a simple, qualitative, but yet accurate numerical simulation, we demonstrated how X-ray polarimetry can be used as a probe independent of 
spectral analyses. As polarization signatures are strongly correlated with the morphology of the coronal plasma, strong constraints can be given by 
future observations.

\subsection{AGN: the test case of MCG-6-30-15}
\label{AGN}

To illustrate the predictive power of X-ray polarimetry in AGN, we focus on the famous candidate MCG-6-30-15 rather than going through a generic model, 
as precise fitting parameters for different spectroscopic models of the object are given in the literature. The Seyfert~1 galaxy is one of the best examples 
showing an asymmetric, broad, 6.4~keV iron line. MCG-6-30-15's red-wing is well established since its \textit{ASCA} discovery by \citet{Tanaka1995} and 
following long observations with {\it XMM-Newton} \citep{Wilms2001,Fabian2002} and {\it Suzaku} \citep{Miniutti2007}. Its 4 --7 keV spectrum is well fitted 
applying the light bending scenario with a rotating SMBH \citep{Wilms2001,Fabian2002}. However, \citet{Inoue2003} and \citet{Miller2008,Miller2009} also 
claimed to be able to reproduce the excess emission above 10~keV, the spectral shape, and the time-invariant red wing of the iron line in MCG-6-30-15 using 
absorption models. In the following, we recall the spectropolarimetric predictions computed by \citet{Marin2012b} according to the prescriptions given by 
the authors of the two competitive models.

\subsubsection{Spectropolarimetric predictions}
\label{AGN:models}

To evaluate the polarization signature resulting from a reflection-dominated \citep{Miniutti2004} versus an absorption-dominated model \citep{Miller2008,Miller2009}, 
we constructed numerical models for MCG-6-30-15 with the characteristics summarized in Tab.~\ref{Tab_AGN}.

\begin{table}
  \begin{center}
    \begin{tabular}{ll}
      \hline
      MCG-6-30-15 (reflection) & MCG-6-30-15 (absorption) \\
      \hline
      Photon index $\alpha$: 1 & Photon index $\alpha$: 1 \\
      Inclination: 30$^\circ$ & Inclination: 30$^\circ$ \\
      Lamp-post height: 2.5~$R_{\rm G}$ & Number of covering media: 2 (zones 4 \& 5) \\
      Spin $a$: 1 & Zone 4: Covering factor: 62~\% ($\tau_{\rm c} \sim 1.5$) \\
      SMBH mass: 1.5 $\times 10^6 \rm M_\odot$ & Zone 5: Covering factor: 17~\% ($\tau_{\rm c} \sim 0.02$) \\
      \hline
    \end{tabular}
    \caption{Parametrization of MCG-6-30-15 model according to 
	     \citet{Miniutti2004} and \citet{Miller2008,Miller2009}.}
    \label{Tab_AGN}
  \end{center}
\end{table}

The reflection scenario consists of a maximally rotating BH with a neutral accretion disk illuminated by an elevated lamp-post, irradiating
an isotropic, unpolarized primary continuum. The input spectrum ranges from 1 to 100 keV and has a power law shape $F_{\rm *}~\propto~\nu^{-\alpha}$ with $\alpha~=~1.0$.
The re-emitted intensity as a function of incident and re-emission angle is computed by the Monte-Carlo radiative transfer code {\it NOAR} \citep{Dumont2000},
the local polarization being estimated according to the transfer equations of \citet{Chandrasekhar1960}. The local, polarized reflection spectra are then 
combined with the {\it KY}-code \citep{Dovciak2004}, that conducts relativistic ray-tracing between the elevated source, the disk, and the distant observer. 
Our choice of parameters is in good agreement with the assumptions of \citet{Miniutti2004}.

The absorption scenario considers a clumpy distribution of Compton-thick, spherical clouds with equal radius and constant density, localized between 1.0 and 1.8 parsecs 
from the irradiating source. The source itself is defined as a geometrically thin, emitting slab that represents the so-called hot inner flow. It irradiates the 
same unpolarized primary spectrum as in the light bending scenario, with the same power-law parameters. In order to avoid confusion with relativistic disk reflection, 
the emitting region is insensitive to scattering and does not reach the ISCO. The model of \citet{Miller2007,Miller2008} uses 5 covering zones to reproduce the 
{\it Chandra} and {\it XMM-Newton} grating data, but only zones 4 and 5 are responsible for the spectral curvature below 10~keV. We therefore focused on these 
two zones and used the latest version of the Monte Carlo code {\sc stokes} \citep{Goosmann2007,Marin2012a} to compute the resulting polarization. Please refer to 
\citet{Marin2012b} for further details about the numerical simulations.

\begin{figure}[!t]
   \begin{center}
   \includegraphics*[width=12cm]{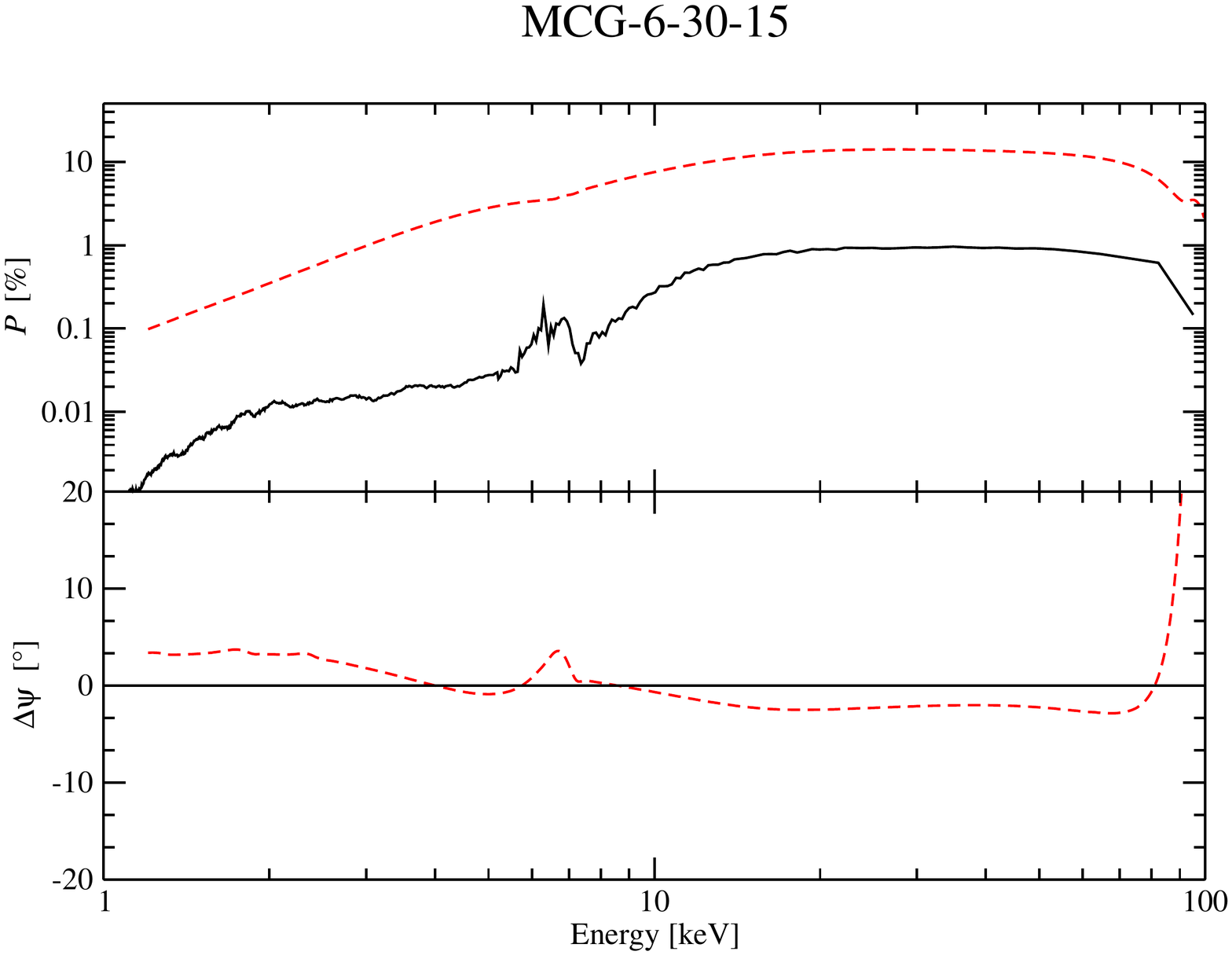}
   \end{center}
   \caption{MCG-6-30-15's percentage of polarization $P$ and variation of the polarization angle 
	  $\Delta\Psi$ with respect to its mean as a function of the energy. 
	  MCG-6-30-15 is viewed at an inclination $i$~=~$60^\circ$ to the axis of symmetry. 
	  $Legend$: a fragmented absorption region (solid line) and a relativistic reflection model 
	  with an extreme Kerr SMBH with $a~=~1$ (red dashed line). The figure is taken from \citet{Marin2012b}.}
   \label{SMBH}
\end{figure}

~\

Fig.~\ref{SMBH} shows the results obtained for MCG-6-30-15, in terms of percentage of polarization $P$ and variation of the polarization position angle $\Delta\Psi$, 
with respect to a convenient average of $\Psi$ over the depicted energy band.

The relativistic $P$ is found to be at least ten times higher than the absorption scenario, with a maximum obtained in the Compton hump, where multiple scattering 
dominates. The spectral shape of $P$ is determined by 1) the net integration of the polarization over the accretion disk in the relativistic case or 2) by the polarization 
phase function of electron scattering in the absorption scenario. In addition to that, the polarization spectra are influenced by dilution from the continuum source, 
whose power-law index $\Gamma$ ($\Gamma$~=~$\alpha$ +1) is set to favor the emission of soft, unpolarized X-ray photons, explaining the diminution of $P$ below 5~keV.

Additional constraints on the origin of iron line come from the variation of the polarization angle. In the relativistic case, $\Delta\Psi$ varies continuously over 
the whole energy band, showing a particularly strong feature across the iron line. The energy-dependent albedo and scattering phase function of the disk material 
explain the smooth behavior of $\Delta\Psi$, which varies by $5^\circ$ around the iron line. The absorption model responds very differently and shows no variation at 
all. Thus, if the $\Delta\Psi$ spectrum of MCG-6-30-15 is found to vary with photon energy, it would be a strong indication for the light bending model to be favored.

\subsubsection{Observational prospects}
\label{AGN:MDP}

We saw from Fig.~\ref{SMBH} that, in principle, X-ray polarization can distinguish between the two scenarios for the origin of the broad iron line in MCG-6-30-15. 
The polarization degree found in the relativistic case is always superior to the one of complex absorption by a factor of ten, with the strongest polarization signal 
to be measured in the 10--100 keV band. Additional constraints, coming from the variation of the polarization angle, could independently give an insight of the more 
probable scenario.

The question so far is: could any X-ray polarimeter have detected such levels of polarization? Knowing that the last measurement of X-ray polarization goes back 
decades ago and that no measurement has ever been taken of MCG-6-30-15, we now investigate the observational prospects for a variety of proposed X-ray missions 
that included a polarimeter.

\begin{figure}[!t]
   \begin{center}
   \includegraphics*[width=12cm]{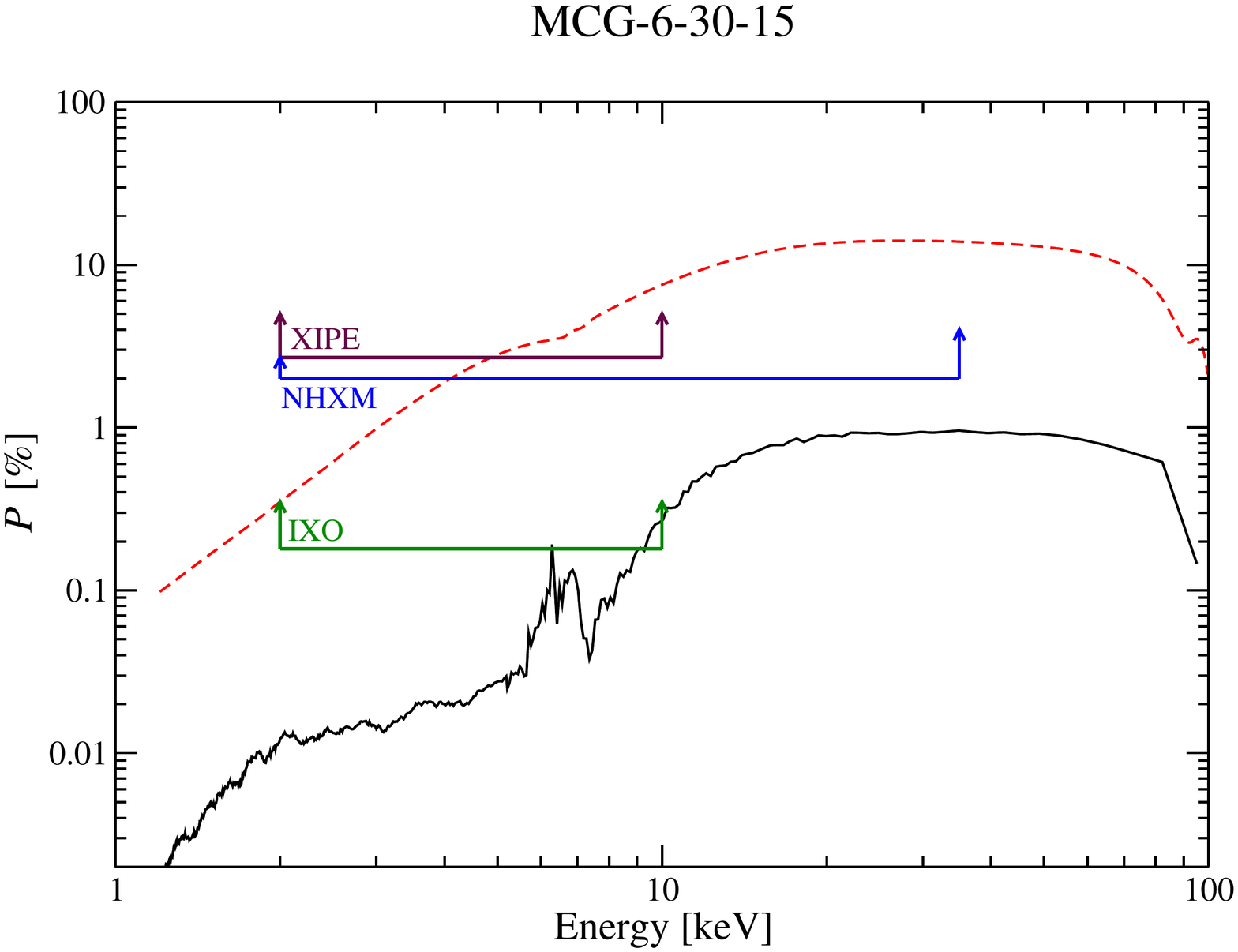}
   \end{center}
   \caption{1~Ms observation minimum detectable polarization of MCG-6-30-15 for the two scenarios of the broad iron line. 
	    The MDP for {\it XIPE} is represented by the maroon line, {\it NHXM} in blue and {\it IXO} in green. 
	    $Legend$: a fragmented absorption region (solid line) and a relativistic reflection model with an extreme 
	    Kerr SMBH with $a~=~1$ (red dashed line).}
   \label{MDP}
\end{figure}

~\

To explore the potential detection of X-ray polarization in MCG-6-30-15, we focused on three mission concepts, namely {\it XIPE} (S-class mission, Soffitta et al. accepted), 
{\it NHXM} (M-class, \citealt{Tagliaferri2012}), and {\it IXO} (L-class, \citealt{NRC2010}). The three polarimeters of these missions were based upon the Gas Pixel Detector 
\citep{Bellazzini2006,Bellazzini2010}, with {\it XIPE} and {\it IXO} instruments ranging from 2 to 10 keV, in comparison with the 2 -- 35 keV observational band of {\it NHXM}. 
The instrumental minimum detectable polarization (MDP) at 99\% confidence level, assuming that MCG-6-30-15's background flux is negligible with respect to the source flux, 
is given by:

\begin{equation}
   \label{MDP_eq}
   {
      MDP~\approx 14\% \left( \frac{S}{1 \rm mCrab} \right)^{-1/2} \left( \frac{\rm exposure \, time}{100,000~\rm s} \right)^{-1/2} 
   }
\end{equation}

and was was computed for a 1~Ms observation with an estimated flux of 2.70~$\pm$~0.15~mCrab in the 2 -- 10 keV band \citep{Krivonos2007}. We took into account the energy dependence
of the detector's modulation factor and calculated the MDP on the full bandpasses of the instrument

The estimated MDP for the two objects are presented in Fig.~\ref{MDP}, using the following color-code : maroon for {\it XIPE}, blue for {\it NHXM} and green for {\it IXO}. 
We find that the reflection scenario of MCG-6-30-15 is within the polarization detectability of {\it XIPE}, so any detection would strongly support the relativistic model. 
Further indications could be deduced from the variation of the polarization position angle if the observed polarization signal is significant enough. The {\it NHXM}'s blue 
line indicates that the polarization signal originating from the relativistic reflection case can be detected almost across the whole 2 -- 35~kev band. It covers, in particular, 
the iron line domain where the $\Delta\Psi$ signature of strong gravity would be detected. Finally, we show that {\it IXO} would have been the only mission to potentially detect 
the low polarization originating from the absorption model. The distinction between the two competitive cases would have come from the detection of either a smooth variation of
$\Psi$ (relativistic signature) or no variation at all (transmission through absorbing gas).

\section{Summary and discussion}
\label{Conclusions}
Conducting detailed radiative transfer simulations that include polarization, we explored an independent and complementary approach to timing and spectral analysis in order 
to probe the origin of the broadening of the iron K$\alpha$ line around stellar and supermassive black holes. 

For the LMXRBs line case, we qualitatively showed that from a spectroscopic point of view, a broad line can be produced by pure Compton scattering mostly with an optically thick corona. 
However the red-shifting, even with the most broadening-efficient geometry of thick corona, is not as large as expected, indicating that the mechanism responsible for the 
distortion of the line has probably to be ascribed to relativistic effects. Nonetheless, in the case of Compton broadening, a significant amount of polarized line flux is 
expected and represents a viable and independent way to discriminate and quantify the different ongoing processes. Moreover, the polarization signal strongly depends on 
the geometry of the scattering material and represents a unique tool to infer it. In our simulations, we demonstrated that the expected polarization signal produced by a 
spherical corona is almost zero, while the axisymmetric slab geometry is more significant in terms of polarization degree ($P \ge$ 10~\%).

Looking at broad iron line signatures in AGN, we recalled the polarization differences emerging from the two main competing scenarios in MCG-6-30-15. The resulting 
polarization signal is found to be rather different between the light bending model and the absorption scenario, with the relativistic reflection showing a polarization 
degree at least ten times higher over the whole energy band. An independent measure of the variation of the photon polarization angle can help to discriminate between 
the two interpretations, as distant absorption should cause no variation of $\Psi$, on the contrary of the smooth and continuous variation of $\Psi$ across the iron 
line in the relativistic case. In this note, we demonstrated that the polarization signal of the light bending model is within the detection range of past X-ray mission 
projects, even for a small, path-finder polarimetry mission. So, any detection of X-ray polarization would strongly support the relativistic scenario.

~\

The lack of broadband spectral observation, covering the full X-ray domain from 0.1~keV to the high energy cut-off, is one of the reasons preventing a final determination 
of the main asymmetrical, broadening mechanism. All the models might not be able to reproduce both the Fe~K$\alpha$ line and the Compton hump, raising a potential way to
disentangle the problem using the future observational campaigns of {\it Astro-H} \citep{Takahashi2010} and {\it NuStar} \citep{Harrison2010}. 
However, one can also reverse the question by looking at the other end of the X-ray domain: \citet{Fabian2009}'s and \citet{Zoghbi2010}'s {\it XMM-Newton} observation 
of the narrow-line Seyfert~1 galaxy 1H0707-495 revealed the presence of a broad feature below 1~keV, assumed to be skewed, fluorescent iron L emission. In the case of 
a high iron abundance in the accretion disk, Fe L emission features should be detectable \citep{Fabian2009} and thus can provide another test for the disambiguation 
between light bending, Compton broadening or complex absorption. Similarly to the case of the Fe~K$\alpha$ line, polarimetric measurements around the Fe L emission 
band could help to favor one of the possible scenarios. 

~\

Encouraging prospects can be foreseen from the recently launched {\it NuStar} satellite. The first {\it NuStar} paper focuses on NGC~1365, 
a nearby galaxy ($z$~=~0.005) hosting a type-1.8 AGN. Simultaneously observed with {\it XMM-Newton}, the broadband spectra (3 -- 79~keV) of NGC~1365 
gave prior to reflection-dominated models using temporal and spectral arguments \citep{Risaliti2013}. However, according to \citet{Miller2013}, the absorption 
model explored by \citet{Risaliti2013} suffers from inaccurate computation. \citet{Miller2013} provided a set of physical parameters associated with a complex model 
of anisotropic cloudlets distribution that should correctly reproduce NGC~1365's spectral shape, excess emission above 10~keV and time-invariant red-wing. NGC~1365 
is rather different from MCG-6-30-15, as it seems to oscillate between type-1 and type-2 classification due to its extreme inclination and rapid X-ray spectral changes, 
indicating the presence of cold gas along the observer's line-of-sight \citep{Risaliti2005}. In this context, a clear spectral or timing disambiguation might not be 
simple. We currently explore the resulting polarization signal induced by the two opposite models (light bending, \citealt{Risaliti2013}, versus complex absorption, 
\citealt{Miller2013}) and preliminary results (Marin et al., submitted to MNRAS) indicate that the two scenarios show even stronger polarization differences than 
in the case of MCG-6-30-15 \citep{Marin2012b}.

\end{document}